\documentclass[prd,
               12pt,tightenlines,
               showpacs,nofootinbib,amsmath,amssymb]{revtex4}

\newcommand{\eV}{\mbox{ \upshape\textrm{eV}}}
\newcommand{\keV}{\mbox{ \upshape\textrm{keV}}}

\newcommand{\GeV}{\mbox{ \upshape\textrm{GeV}}}
\newcommand{\meV}{\mbox{ \upshape\textrm{meV}}}
\newcommand{\abs}[1]{\mbox{$\left| #1 \right|$}}
\newcommand{\mbb}{m_{\beta\beta}}
\newcommand{\e}{\mathrm{e}}

\begin{document}

\title{$\nu$MSM---Predictions for Neutrinoless Double Beta Decay}
\author{F. Bezrukov}
\email{fedor@ms2.inr.ac.ru}
\affiliation{
    Institute for Nuclear Research of the Russian Academy of Sciences,\\
    60th October Anniversary prospect 7a, Moscow 117312, Russia
}
\date{May 30, 2005}

\begin{abstract}
  We give the prediction on the effective Majorana mass for
  neutrinoless double $\beta$ decay in a simple extension of the
  Standard Model ($\nu$MSM).  The model adds three right-handed
  neutrinos with masses smaller than the electroweak scale, and
  explains dark matter of the Universe.  This leads to constraints
  $1.3\meV<\mbb^{NH}<3.4\meV$ in normal neutrino mass hierarchy and
  $13\meV<\mbb^{IH}<50\meV$ in inverted hierarchy.
\end{abstract}

\maketitle

\section{Introduction}

Currently most laboratory experiments are described up to a very good
precision by the Standard Model of particle interactions.  However,
recent developments show that effects beyond the Standard Model surely
exist.  The anomaly in atmospheric neutrinos is now understood by
$\nu_\mu \rightarrow \nu_\tau$
oscillation~\cite{Ashie:2005ik,Ashie:2004mr}, while the solar neutrino
puzzle is solved by the oscillation $\nu_e \rightarrow \nu_{\mu,
\tau}$~\cite{Ahmed:2003kj,Araki:2004mb} incorporating the MSW LMA
solution
\cite{Wolfenstein:1977ue,Mikheev:1986wj,Mikheev:1986if,Mikheev:1986gs}.
Current data are consistent with flavor oscillations between three
active neutrinos\footnote{We do not include the LSND
anomaly~\cite{Aguilar:2001ty} in present analysis.} with parameters
given in table~\ref{tab:noscdat}.  The definition of mixing angles is
usual
\begin{equation}\label{e:u3}
  \left(\begin{matrix}
    \nu_e \\ \nu_{\mu} \\ \nu_{\tau}
  \end{matrix}\right)
  =
  \left(\begin{matrix}
    c_{12}c_{13} & s_{12}c_{13} & s_{13}
    \\
    -s_{12}c_{23}-c_{12}s_{23}s_{13}\e^{i\delta} &
      c_{12}c_{23}-s_{12}s_{23}s_{13}\e^{i\delta} &
      s_{23}c_{13} \e^{i\delta}
    \\
    s_{12}s_{23}-c_{12}c_{23}s_{13}\e^{i\delta} &
      -c_{12}s_{23}-s_{12}c_{23}s_{13}\e^{i\delta} &
      c_{23}c_{13} \e^{i\delta}
  \end{matrix}\right)
  \times \left(\begin{matrix}
    \e^{i\alpha_1/2}~\nu_1 \\ \e^{i\alpha_2/2}~\nu_2 \\ \nu_3
  \end{matrix}\right)
  \;,
\end{equation}
where $s_{ij}\equiv\sin\theta_{ij}$, $c_{ij}\equiv\cos\theta_{ij}$,
$\delta$ is the usual CP-violating phase and $\alpha_1$, $\alpha_2$
are Majorana phases.  The three neutrino masses $m_i$ should be added
to the parameter set that describes the matrix (\ref{e:u3}),
representing therefore nine unknown parameters altogether.

\begin{table}
  \caption{Neutrino oscillation parameters (2004 status)}
  \label{tab:noscdat}
  \begin{center}
    \begin{tabular}{ccl}
      Parameter & Value $\pm1\sigma$ & Comment \\
      \hline
      $\Delta m_{21}^2$     & $7.9_{-0.5}^{+0.6}\times10^{-5}\eV^2$
        & Solar $\nu$ \cite{Araki:2004mb,Ahmed:2003kj}\\
      \hline
      $\tan^2\theta_{12}$         & $0.40_{-0.07}^{+0.10}$
        & For $\theta_{13}=0$ \cite{Araki:2004mb,Ahmed:2003kj} \\
      \hline
      $|\Delta m_{32}^2|$   & $2.0_{-0.4}^{+0.6}\times10^{-3}\eV^2$
        & Atmospheric $\nu$ \cite{Ashie:2005ik} \\
      \hline
      $\sin^2 2\theta_{23}$ & $>0.95$
        & For $\theta_{13}=0$ \cite{Ashie:2005ik}\\
      \hline
      $\sin^2 \theta_{13}$ & $<0.016$
        & \cite{Bahcall:2004ut} \\
      \hline
    \end{tabular}
  \end{center}
\end{table}

Another strong indication that the Standard Model is not complete
comes from cosmology.  Recently, various cosmological observations
have revealed that the universe is almost spatially flat and mainly
composed of dark energy ($\Omega_{\Lambda}=0.73 \pm 0.04$), dark
matter ($\Omega_\mathrm{dm} = 0.22 \pm 0.04$) and baryons ($\Omega_b =
0.044 \pm 0.004$)~\cite{Eidelman:2004wy}.

A promising way of extending the Minimal Standard Model leading to
explanation of these facts was proposed in~\cite{Asaka:2005an}.  The
idea is to add 3 right-handed neutrinos to the model with the most
general gauge-invariant and renormalizable Lagrangian.  One then
requires that active neutrinos satisfy the known oscillation data, and
the (Warm) Dark
Matter~\cite{Peebles:1982ib,Olive:1981ak,Dodelson:1993je,Shi:1998km,%
Dolgov:2000ew,Abazajian:2001nj} %
is given by the right-handed (sterile) neutrinos (one could also try
to add only 2 right-handed neutrinos, what is enough to explain
oscillation data, but it turns out to be inconsistent with sterile
neutrino being Dark Matter).  These surprisingly leads to a stringent
constraint on the active neutrino masses---the lightest neutrino
should have the mass less than about $10^{-5}\eV$.

Baryon number asymmetry of the Universe can also be explained in
$\nu$MSM, see Ref.~\cite{Asaka:2005pn}.  More constraints on the
parameters of the sterile neutrinos appear from that consideration,
but no additional restrictions on active neutrino parameters, that are
relevant for the current discussion are introduced.

We are going to analyze the effective Majorana mass for neutrinoless
double beta decay emerging in this model.  Section~\ref{sec:nuMSM}
revives the main points of $\nu$MSM relevant for our discussion, and
section~\ref{sec:0nubb} provides the estimate of the effective
Majorana mass for neutrinoless double $\beta$ decay in the model.

\section{The $\nu$MSM Model}
\label{sec:nuMSM}

Lagrangian of $\nu$MSM, introduced in~\cite{Asaka:2005an} adds 3 right
handed neutrinos to the Standard Model, which are SU(2)$\times$U(1)
singlets and have the most general gauge-invariant and renormalizable
interactions:
\begin{eqnarray*}
  \delta {\cal L}
  = \overline{N_I} i \partial_\mu \gamma^\mu N_I
  - f^{\nu}_{I\alpha} \, \Phi  \overline{N_I} L_\alpha
  - \frac{M_I}{2} \; \overline{N_I^c} N_I + h.c. \,,
\end{eqnarray*}
where $\Phi$ and $L_\alpha$ ($\alpha=e,\mu,\tau$) are the Higgs and
lepton doublets, respectively, and both Dirac ($M^D = f^\nu \langle
\Phi \rangle$) and Majorana ($M_I$) masses for neutrinos are
introduced. We have taken a basis in which mass matrices of charged
leptons and right-handed neutrinos are real and diagonal.
In~\cite{Asaka:2005an} this model was called ``the \emph{$\nu$ Minimal
Standard Model} (the $\nu$MSM)''.

Let us first discuss neutrino masses and mixing in the $\nu$MSM.  We
will restrict ourselves to the region in which the Majorana neutrino
masses are larger than the Dirac masses, so that the seesaw
mechanism~\cite{Yanagida:1980xy} can be applied.  Note that this does
not reduce generality since the latter situation automatically appears
when we require the sterile neutrinos to play a role of dark matter,
as we shall see. Then, right-handed neutrinos $N_I$ become
approximately the mass eigenstates with $M_1 \le M_2 \le M_3$, while
other eigenstates can be found by diagonalizing the mass matrix:
\begin{eqnarray*}
  M^\nu = \left(M^D\right)^T \; M_I^{-1} \; M^D \,.
\end{eqnarray*}
which we call the seesaw matrix.  The mass eigenstates $\nu_i$
($i=1,2,3$) are found from
\begin{eqnarray*}
  U^T M^\nu U = M^\nu_\mathrm{diag} =
  \mathrm{diag}(m_1, m_2, m_3 ) \,,
\end{eqnarray*}
and the mixing in the charged current is expressed by
$\nu_\alpha = U_{\alpha i} \, \nu_i + \Theta_{\alpha I}\, N_I^c$
where $\Theta_{\alpha I} = (M^D)^\dagger_{\alpha I} M_I^{-1} \ll
1$ under our assumption.  This is the reason why right-handed
neutrinos $N_I$ are often called ``sterile'' while $\nu_i$
``active''.

For three sterile neutrinos added to the SM all active neutrinos
acquire masses, and the smallest mass can be in the range $0 \le
m_\mathrm{min} \lesssim \mathcal{O}(0.1)\eV$~\cite{Seljak:2004xh}.  In
particular, the degenerate mass spectra of active neutrinos are
possible when $m_\mathrm{min}^2 \gtrsim \Delta m_\mathrm{atm}^2$.
Note also that there are two possible hierarchies in the masses of
active neutrinos, i.e.\ ``normal'' hierarchy $\Delta m_{32}^2>0$
leading to $m_1<m_2<m_3$, and ``inverted'' hierarchy $\Delta
m_{32}^2<0$ with $m_3<m_1<m_2$.  Note, that here $\nu_1$ is the mass
state maximally mixed with the electron flavor neutrino and $\nu_3$ is
the mass state maximally mixed with $\tau$ neutrino (this is different
from the convention $m_1<m_2<m_3$ used in \cite{Asaka:2005an}).

When the active-sterile neutrino mixing $\abs{\Theta_{\alpha I}}$ is
sufficiently small, the sterile neutrino $N_I$ has never been in
thermal equilibrium and is produced in non-equilibrium reactions.  The
production processes include various particle decays and conversions
of active into sterile neutrinos.  Requirement that enough of Dark
Matter neutrino is produced leads to the following constraint on the
Dirac mass term in the Lagrangian (see Ref.~\cite{Asaka:2005an})
\begin{equation}
  \label{eq:DMCondition}
  \sum_I \sum_{\alpha = e,\mu,\tau} \abs{ M^D_{I \alpha}}^2
  = m_0^2 \,,
\end{equation}
where $m_0 = {\cal O}(0.1)\eV$ and the summation over $I$ is only over
sterile neutrinos being Warm Dark Matter.  Notice that this constraint
on dark-matter sterile neutrinos is independent of their masses, at
least for $M_I$ in the range discussed below.

The sterile neutrino, being warm dark matter, further receives
constraints from various cosmological observations and the
possible mass range is very restricted as
\begin{equation*}
  2 \keV \lesssim M_I \lesssim 5 \keV \,,
\end{equation*}
where the lower bound comes from the cosmic microwave background and
the matter power spectrum inferred from Lyman-$\alpha$ forest
data~\cite{Viel:2005qj}, while the upper bound is given by the
radiative decays of sterile neutrinos in dark matter halos limited by
X-ray observations~\cite{Abazajian:2001vt}.

The constraint~(\ref{eq:DMCondition}) together with the neutrino
oscillation data leads to the following conclusion.  At least $3$
right-handed neutrinos are required.  In case of only 3 sterile
neutrinos only one of them can play the role of WDM (let it be $M_1$,
for definiteness), and the mass of the lightest active neutrino
$m_\mathrm{min}$ should be less than about $10^{-5}\eV$ (see
Ref.~\cite{Asaka:2005an} for details).  If there are more than three
sterile neutrinos no constraint is present.

In the work~\cite{Asaka:2005pn} baryon asymmetry of the Universe was also
explained in the framework of $\nu$MSM.  Additional constraints from
requirement of correct baryon asymmetry arise on the parameters of the
sterile neutrinos in $\nu$MSM, but no additional constraints appear on
the parameters of the active neutrinos relevant for our discussion
here.

\section{Neutrinoless Double Beta Decay Effective Mass}
\label{sec:0nubb}

The constraints described in the previous section allow to determine
the effective mass for neutrinoless double beta decay.  This mass is
related to the mass eigenvalues and mixings by
\begin{equation}\label{mbb}
  \mbb=\left|
    \sum_i m_iU_{ei}^2
    +M_1\Theta_{e1}^2
  \right|
  \;,
\end{equation}
where the first term corresponds to the standard three neutrino
contribution, and the second one is the contribution from the Dark
Matter sterile neutrino.  The other two sterile neutrinos are
considered heavy ($\gtrsim10\GeV$, see Ref.~\cite{Asaka:2005pn}) and
do not contribute to $\mbb$.

First, let us estimate the contribution from the last term.  Using
definition of $\Theta_{e1}$ we get
\[
  |M_1\Theta_{e1}^2| = \frac{|{M^{D}_{1e}}^2|}{M_1}
  \;.
\]
The dark matter constraint~(\ref{eq:DMCondition}) requires
$|{M^{D}_{1e}}^2|\lesssim (0.1\eV)^2$.  So, the absolute value of the whole
contribution is (as far as $M_1\simeq O(1)\keV$)
\[
  |M_1\Theta_{e1}^2| < 10^{-5}\eV
  \;.
\]
This means that it can be neglected for any reasonable contribution
from the first term in~(\ref{mbb}).

So, standard analysis of the formula~(\ref{mbb}) can be applied (see
eg.~\cite{Aalseth:2004hb}).  For normal neutrino mass hierarchy we have
(as far as $m_1$ can be neglected)
\begin{equation*}
  \mbb^{NH} = \left|
    \sqrt{\Delta m_{21}^2}\sin^2\theta_{12}\cos^2{\theta_{13}}
    +\sqrt{|\Delta m_{31}^2|}\sin^2\theta_{13}\e^{-i\alpha_2}
  \right|
  \;.
\end{equation*}
For $\theta_{13}=0$ this leads to $\mbb^{NH}=2.6\pm0.4\meV$.  Using
$1\sigma$ bound on $\theta_{13}$ from \cite{Bahcall:2004ut}, we get
$1.3\meV<\mbb^{NH}<3.4\meV$.  It is worth noting, however, that for
$\tan^2_{13}\ge\sin^2\theta_{12}\sqrt{|\Delta m_{21}^2|/\Delta
m_{31}^2}\sim 0.06$ complete cancellation may occur, so at $3\sigma$
level $\mbb^{NH}$ can be zero.

In the case of inverted hierarchy, neglecting $m_3$, one obtains
\begin{equation*}
  \mbb^{IH} =\sqrt{|\Delta m_{31}^2|}
    \cos^2\theta_{13}
    \sqrt{1-\sin^22\theta_{12}\sin^2\frac{\alpha_2-\alpha_1}{2}}
  \;.
\end{equation*}
So, we get $13\meV<\mbb^{IH}<50\meV$.

\section{Conclusions}

In the $\nu$MSM model the lightest active neutrino has the mass
$<10^{-5}\eV$, and there is a relatively light sterile neutrino with
the mass $2\keV\lesssim M_1\lesssim5\keV$ and mixing with active
neutrinos of the order of $10^{-4}$, which plays a role of the Warm
Dark Matter.  Though it is quite light, the sterile dark matter
neutrino makes a vanishing contribution to effective neutrinoless
double $\beta$ decay Majorana mass $m_{\beta\beta}$ because of its
small mixing angle.  Thus, predictions for $m_{\beta\beta}$ can be
obtained from usual analysis with zero lightest neutrino mass.
Specifically, current $1\sigma$ limits are
\[
  1.3\meV<\mbb^{NH}<3.4\meV
\]
for normal active neutrino mass hierarchy and
\[
  13\meV<\mbb^{IH}<50\meV
\]
for inverted hierarchy.

\begin{acknowledgments}
Author is indebted to Mikhail Shaposhnikov for drawing his interest to
$\nu$MSM, numerous invaluable discussions of the properties of the
model and inspiration for the work.  The work of F.B. is supported in
part by INTAS YSF 03-55-2201 and Russian Science Support Foundation.
\end{acknowledgments}


\end{document}